\documentclass{article}
\usepackage{ijcai23}

\usepackage{times}
\usepackage{soul}
\usepackage{url}
\usepackage[hidelinks]{hyperref}
\usepackage[utf8]{inputenc}
\usepackage[small]{caption}
\usepackage{graphicx}
\usepackage{amsmath}
\usepackage{amsthm}
\usepackage{booktabs}
\usepackage{algorithm}
\usepackage{algorithmic}
\usepackage[switch]{lineno}
\usepackage{subfigure}
\usepackage{multirow}
\usepackage{cleveref}

\crefformat{section}{\S#2#1#3} 
\crefformat{subsection}{\S#2#1#3}
\crefformat{subsubsection}{\S#2#1#3}

\title{Re-creation of Creations: A New Paradigm for Lyric-to-Melody Generation}

\author{
Ang Lv$^1$
\and
Xu Tan$^2$\and
Tao Qin$^2$\and
Tie-Yan Liu$^2$ \And
Rui Yan$^1$
\affiliations
$^1$Renmin University of China\\
$^2$Microsoft Research Asia
\emails
\{anglv, ruiyan\}@ruc.edu.cn,
\{xuta, taoqin, tyliu\}@microsoft.com,
}

\begin{document}

\maketitle

\begin{abstract}
Lyric-to-melody generation is an important task in songwriting, and is also quite challenging due to its unique characteristics: the generated melodies should not only follow good musical patterns, but also align with features in lyrics such as rhythms and structures. These characteristics cannot be well handled by neural generation models that learn lyric-to-melody mapping in an end-to-end way, due to several issues: (1) lack of aligned lyric-melody training data to sufficiently learn lyric-melody feature alignment; (2) lack of controllability in generation to better and explicitly align the lyric-melody features. In this paper, we propose Re-creation of Creations (ROC), a new paradigm for lyric-to-melody generation. ROC generates melodies according to given lyrics and also conditions on user-designated chord progression. It addresses the above issues through a generation-retrieval pipeline. Specifically, our paradigm has two stages: (1) creation stage, where a huge amount of music fragments generated by a neural melody language model are indexed in a database through several key features (e.g., chords, tonality, rhythm, and structural information); (2) re-creation stage, where melodies are re-created by retrieving music fragments from the database according to the key features from lyrics and concatenating best music fragments based on composition guidelines and melody language model scores. ROC has several advantages: (1) It only needs unpaired melody data to train melody language model, instead of paired lyric-melody data in previous models. (2) It achieves good lyric-melody feature alignment in lyric-to-melody generation. Tested by English and Chinese lyrics, ROC outperforms previous neural based lyric-to-melody generation models on both objective and subjective metrics.
Demos are in \url{https://ai-muzic.github.io/roc}, and the code
is in \url{https://github.com/microsoft/muzic/}.
\end{abstract}

\section{Introduction}
In recent years, with the development of artificial intelligence, researchers have achieved great success in various aspects of automatic songwriting such as lyric generation~\cite{10.1145/2939672.2939679,DBLP:journals/corr/abs-2107-01875}, melody generation~\cite{WU2020103303,xiaoice}, lyric-to-melody generation~\cite{DBLP:journals/corr/abs-1809-04318,10.1145/3424116,DBLP:journals/corr/abs-2012-05168,DBLP:journals/corr/abs-2109-09617}, and melody-to-lyric generation~\cite{10.1145/3474085.3475502,DBLP:journals/corr/abs-2107-01875,li-etal-2020-rigid}. Among all directions, lyric-to-melody generation is one of the most fundamental tasks and is the focus of this paper. A high-quality lyric-to-melody generation should not only focus on beautiful melodies but also align rhythms and structures in lyrics with melodies.

Among all methods in lyric-to-melody generation, rule-based and neural-based are two main categories.  Rule-based methods incorporate composition guidelines summarized by composers so that the rhythm and structure alignment between lyrics and melodies can be basically ensured. However, they require too much labor and music expertise. Currently, the end-to-end neural generation model is the mainstream method but it also suffers from many weaknesses. First, the mapping from lyric to melody is hard to learn because the melody is weakly correlated with the lyric (e.g., a melody can be accompanied by different lyrics as long as the lyric syllables align with notes). Therefore, a large amount of aligned training data, which is rare however, is required. Second, an end-to-end model is a black box with a weak guarantee of lyric-melody feature alignment, resulting in low-quality of generation. SongMASS~\cite{DBLP:journals/corr/abs-2012-05168} proposes an unsupervised method to train lyric-to-lyric and melody-to-melody models respectively, and learns the alignment between two models. It sidesteps the insufficiency of paired data but still suffers from insufficient feature alignment between lyrics and melodies. TeleMelody~\cite{DBLP:journals/corr/abs-2109-09617} proposes a two-stage generation: lyric-to-template and template-to-melody. With templates, it does not need paired data, and some composition guidelines for tonality, rhythm and chord progression are better considered. However, features still can not be aligned explicitly because two stages are both neural-based, and also there are error accumulations.

Considering above weaknesses in existing methods, we combine merits of both rule-based and neural-based methods, and propose Re-creation of Creations (ROC), a new paradigm for conditional lyric-to-melody generation with a generation-retrieval pipeline. There are two stages in ROC: creation and re-creation. 
In creation stage, we use melody data to train a melody language model, which is used to generate high-quality music fragments. Generated melody fragments are stored in a database indexed by extracted key features including tonality, rhythm, chord, structural information, e.g. belonging to a chorus or a verse. These features are used as the key for retrieval in the next stage. 
In re-creation stage, ROC retrieves melody fragments from the database for lyrics (sentence by sentence) and concatenates them to compose a melody, conditioned on user-designated chord progression and tonality. 
In detail, we first infer the features of a lyric sentence where these features are used as the query for retrieval, and for which we propose a lyric structure recognition algorithm to extract structure features by segmenting lyrics into choruses and verses. 
Then, we retrieve melody fragments by matching the features of a lyric sentence to those of melody fragments, and re-rank retrieved melody fragments by both composition guidelines and melody language model scores. 
To better align the rhythm and structure in melodies and lyrics, we propose a melody-sharing scheme between lyrics with similar rhythm patterns (e.g., sharing among chorus or verses).
The best melody fragments for each lyric are concatenated as a complete song which is then polished because concatenation may cause some issues like overlapping bars. ROC has the following advantages. (1) ROC does not need paired lyric-melody data because we only train a melody language model. (2) The non end-to-end pipeline, retrieval-based composition, and composition guidelines in ROC can better align rhythm and structure between lyrics and melodies than previous works.

To sum up, our main contributions are as follows:

(1) We propose ROC, a new paradigm for conditional lyric-to-melody generation with creation stage and re-creation stage, which does not need paired lyric-melody data for training, and can better align the rhythm and structure between lyrics and melodies. 

(2) We make a series of deigns to ensure ROC work effectively, including a lyric structure recognition algorithm, a short melody fragment generation procedure, a retrieval and re-ranking procedure, melody sharing scheme, and melody polishing, etc.

(3) Experimental results demonstrate that ROC outperforms end-to-end and non end-to-end baselines on both objective and subjective metrics.

\section{Background}
\subsection{Characteristics of Melodic Songs}
Empirically, beautiful and harmonic songs in pop music have common characteristics in lyric, melody, and lyric-melody feature alignment. We list a few characteristics that ROC take advantage of in below. We omit lyric characteristics because we only consider how to generate melodies from given lyrics in lyric-to-melody generation task.

\subsubsection{Melody}
\label{sec:2-1-1}
The following melody patterns are crucial to the quality of a song according to some composition guidelines.
\begin{itemize}
    \item Chord progression. A good chord progression can guide the emotion development of a melody. Besides, the chord progression should return to the tonic chord to create a sense of stable and smooth ending.
    \item Tonality. It has an impact on emotional atmosphere. For example, a major sounds enthusiastic, gorgeous, bright and cheerful while a minor sounds cold, melancholy and magical. 
    \item Varied pitch and note density. Average pitch and note density usually increase in choruses for more intensive emotion expression. 
    \item Pitch range. Pitches in the beginning of a song should be mild to make room for lifting in the chorus. Also, pitches usually do not fluctuate too much in a chorus or a verse.
    \item Tendency. Some notes tend to be followed by some specific notes due to the tendency between notes.\footnote{\url{http://www.musicnovatory.com/cqtendency.html}}
\end{itemize}

\subsubsection{Lyric-Melody Feature Alignment.}
\label{sec:2-1-2}
In lyric-to-melody generation, melody should not only follow good musical patterns, but also align with lyrics in some aspects:
\begin{itemize}
    \item The structure of lyrics and melodies should match. In melody, choruses are usually more intensive and reach the climax of the whole song. In lyric, chorus lyrics are usually more lyrical than those in verses. The better match promotes emotion expression.
    \item Lyric segments with the similar rhythm patterns usually share melodies. In a song, one of the most obvious rhythm patterns is that choruses and verses repeat many times. Most lyrics share the same sentence pattern (e.g., the same number of syllables) with their counterparts in other choruses (or verses) and these lyrics usually share the similar melody.
    \item Composers should choose the proper tonality according to lyric sentiments to express emotion thoroughly.
    \item Melody cadences and lyric endings should match for better rhythm and structure alignment.
\end{itemize}

\subsection{Lyric-to-Melody Generation}

\label{sec:related}
In the early days, there are some statistical and rule-based methods for lyric-to-melody generation. ~\cite{Long2013TMusicAM} focus on lyric-note correlation and propose a probability model but ignore music knowledge. ~\cite{japan} study the Japanese prosody and its role in composition and propose a probability model to generate melody. They incorporate more musical patterns but still ignore structural features so that no similar segments are repeating in generated songs which makes it sound not like human-composed. Besides, theses traditional methods require too much labor and expertise of music or linguistics, and thus the research focus turns to neural-based methods.

With the advent of the neural network era, major breakthroughs have been made in the field of lyric-to-melody generation. Many methods~\cite{DBLP:journals/corr/abs-1809-04318,10.1145/3424116,DBLP:journals/corr/abs-2012-05168,DBLP:journals/corr/abs-2109-09617} regards lyric-to-melody generation as a sequence-to-sequence task, which learns a mapping from lyric sentence to melody phrase. Such end-to-end models require a large amount of paired lyric and melody data but insufficient aligned data greatly hinders research. SongMASS ~\cite{DBLP:journals/corr/abs-2012-05168} trains lyric-to-lyric and melody-to-melody models separately then conducts interaction between models to sidestep the lack of paired data. However, it is an end-to-end method and suffers low controllability which results in no guarantee of aligned features between lyric and melody. To be more controllable, TeleMelody~\cite{DBLP:journals/corr/abs-2109-09617} divides the end-to-end generation pipeline into two stages: lyric-to-template and template-to-melody. Templates bridge the gap between lyric and melody. Besides, templates make the generation more controllable, which increases the ease of aligning features between lyrics and melodies. However, two stages in TeleMelody are both neural-based so the alignment can not be ensured explicitly and there are error accumulations that hurt generation quality.

\begin{figure*}[t]
    \centering
    \includegraphics[width=\linewidth]{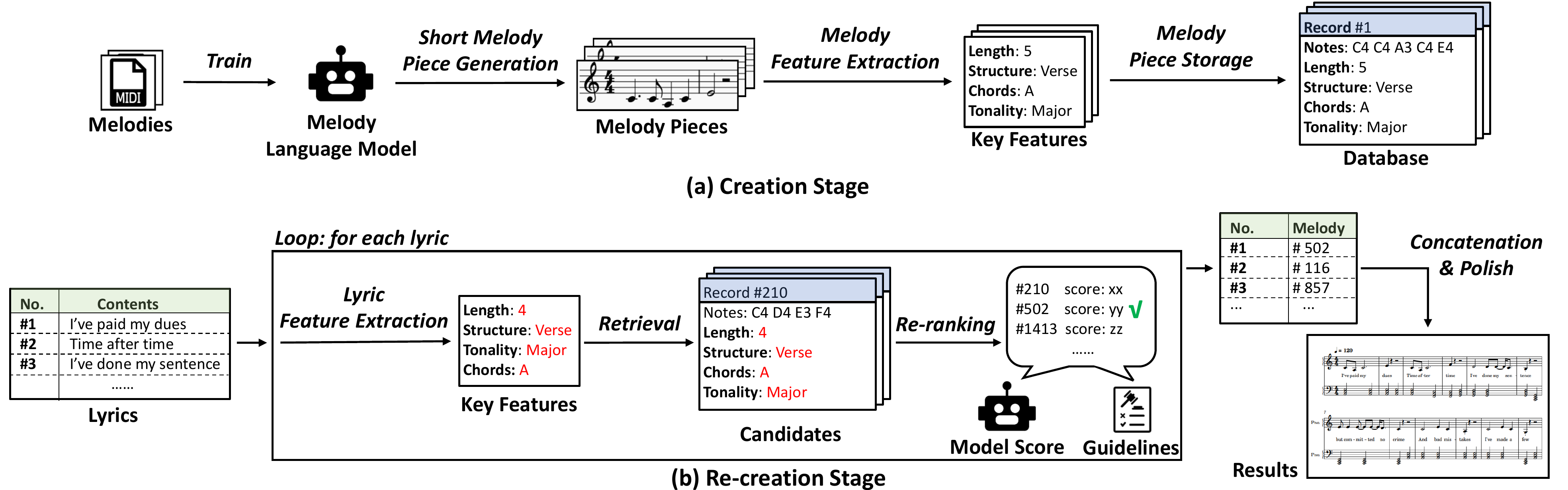}
    \caption{The pipeline of ROC. In creation stage, melody language model generates melody fragments which are stored in the database along with key features. In re-creation stage, we infer the features of each lyric in a song, which are used as the query to retrieve melody fragments. Retrieved fragments are re-ranked by both composition guidelines and the melody language model scores. After concatenating and polishing these melody fragments, ROC generates the final melodies.}
    \label{fig:pipeline}
\end{figure*}

We propose ROC to address aforementioned weaknesses. ROC does not need paired data because only melodies involved in training. In re-creation stage, retrieval and matching enable ROC to explicitly consider lyric and melody features, leading to better feature alignment between lyrics and melodies. We also incorporate composition guidelines to guide the design of ROC to make generation sounds like human-composed.

\section{Methodology}
\label{sec:method}
The whole pipeline of ROC is shown in Figure ~\ref{fig:pipeline}. In ROC, there are two stages: creation and re-creation. Two stages are conducted sequentially. We detail each stage in this section.

\subsection{Creation Stage}

Creation stage prepares the model and data that the next stage use. As shown in Figure ~\ref{fig:pipeline}(a), in creation stage, we use melody data to train a melody language model and let the model generate short melody fragments which are stored in the database and indexed by key features. We introduce the melody language model, melody feature extraction and details about melody fragment storage in below.

\subsubsection{Melody Language Model} 
\label{sec:3-1-1}
To ensure the originality of the generated songs and avoid infringement, we train an auto-regressive melody language model based on transformer architecture with melody data to produce new melody fragments. As an additional benefit, the trained melody language model is reused for re-ranking fragments in re-creation stage, which is discussed in \cref{sec:retrieval}. However, generated songs are not competitive with the training data in quality, due to the neural model's poor ability to generate long sequences. Therefore, we apply a short melody fragment generation procedure: given two melody bars, the trained melody language model only generates the next two bars. The prediction interval of two-bar comes from trials and is an appropriate choice because if the interval is too long, the quality of generated melody fragments is low and re-creation will be inflexible; if the prediction interval is shorter than one bar, the later concatenation and polish in re-creation stage will be too complicated. Predicted fragments are removed if they are the same as their ground truth. In this way, we obtain original melody fragments of high quality.

\subsubsection{Melody Feature Extraction} 
\label{sec:features}
~\cref{sec:2-1-1} describes the importance of lyric-melody structure alignment, chord progression, and tonality, and based on this we summarize four key features in a melody fragment that can be used as keys for storage and retrieval, which we name as `Length', `Structure', `Chords', `Tonality'.
\begin{itemize}
    \item \textbf{`Length'.} It is the number of notes in a melody fragment. We use this feature to basically align the rhythm between lyrics and melodies: In most cases, we decide the retrieval length according to the number of syllables in a lyric and align one syllable with one note. Occasionally, we allow one syllable to align with multiple notes and details are in \cref{sec:retrieval}.
    \item \textbf{`Structure'.} It indicates the fragment belongs to whether a chorus or a verse. In re-creation stage, we also recognize this feature of a lyric for structure alignment. `Structure' is inferred by an algorithm based on self-similarity matrix \cite{pychorus}. 
    \item \textbf{`Chords'.} It is the corresponding chords of the melody fragment. `Chords' is inferred based on note pitch distribution based on a viterbi algorithm \cite{viterbi}.
    \item \textbf{`Tonality'.} It implies the tonality of the melody fragment and can be inferred by \cite{DBLP:journals/corr/abs-2010-08091}. Appropriate tonality expresses the emotion of lyrics and plays a crucial role in re-creation stage: a consistent tonality through the whole song is given based on lyric sentiment, and fragments with unmatched tonality are filtered during retrieval.
\end{itemize}

In Figure ~\ref{fig:pipeline}, we briefly show the appearance of a record. Due to limited pages, we visualize a melody fragment in the database along with its key features in ~\ref{apx:2}, for better understanding our data structure.

\subsubsection{Melody Fragments Storage} 
\label{sec:3-1-3}
A generated two-bar fragment is stored as two one-bar fragments and one two-bar fragment. We ignore the bar index and focus on melodic notes and key features as Figure~\ref{fig:pipeline}(a) shows. Also, we deduplicate melody fragments and filter monotonous fragments. We call a melody fragment as monotonous if there are too few unique pitches in a melody fragment.

\subsection{Re-creation Stage}
Based on the creation stage, ROC composes melodies for lyrics sentence by sentence in re-creation stage. Figure ~\ref{fig:pipeline}(b) illustrates the overview. In this stage, users designate a preferred chord progression to guide the composition and provide lyrics to ROC. Then, ROC infer features (e.g., length, structure, and tonality, etc.) from lyrics and use these features to retrieve melody fragments. Melody candidates are first filtered by composition guidelines and then re-ranked by the melody language model scores. When each lyric in a song has been assigned with melody fragments, we concatenate melody fragments together and polish the song. In this section, we first introduce how to extract features in lyrics as queries for retrieval. Then, we discuss retrieval and re-ranking details. At last, we talk about polish, a post-process to further improve the quality of melodies.

\subsubsection{Lyric Feature Extraction}
\label{sec:lyrics-structure}
As mentioned in ~\cref{sec:2-1-2}, lyric-melody feature alignment matters. Given a lyric, we extract features as queries to retrieve matched melody candidates. Among features mentioned in ~\cref{sec:features}, the `Chords' is inferred based on the chord progression provided by users. `Length' is the number of syllables in a lyric. `Tonality' is automatically set as major or minor based on positive or negative sentiments of lyrics. We use third-party libraries for Chinese \cite{cnsenti} and English \cite{textblob} sentiment analysis. In terms of `Structure', we design a heuristic algorithm to recognize structural information in lyrics and introduce details in below. 

\begin{algorithm}[!t]
\caption{Lyric Structure Recognition with $(K,L)$ $Repeat$ Algorithm.}\label{alg:lyric-recog}
\begin{algorithmic}[1]
\STATE \textbf{Input}:\\
    The string $S$ abstracted from lyrics;\\
    The segmentation granularity $g$.
\STATE \textbf{Initialize}:\\
     Set all elements in $struct$ array as 0.

\WHILE {$True$}
\STATE {Find $R[L,K]$ with the largest $L$ from $S$.}

\IF {$L$ \textgreater $g$ and $K$ \textgreater 1}
    \STATE {Assign the $struct$ value of each element in $R[L,K]_{i}$ as the index in $S$ of each element in $R[L,K]_{1}$, where i $\in$ [2,K].}
    \STATE {Remove elements with non-zero $struct$ value from $S$.}
\ELSE
    \STATE break
\ENDIF
\ENDWHILE
\end{algorithmic}
\end{algorithm}

Recap that lyric segments with similar rhythm patterns should share melodies (~\cref{sec:2-1-2}). To fuse this characteristic into generation, we design an algorithm for searching lyric segments that bear similar sentence patterns. 

First, we define some preliminaries. Assume a song contains $n$ sentences. We represent a sentence with the number of syllables in it. Therefore, lyrics of a song can be abstracted into a number string $S$. We call a substring in $S$ as $(K,L)$ \textit{Repeat} if it is of length $L$ and repeats $K$ times non-overlappingly in $S$. The collection of these repetitive substrings is denoted as $R[L,K]$. $R[L,K]_{i}$ denotes the $i$-th repeat in $R[L,K]$, where i $\in$ [1,K]. Each segment in $R[L,K]$ should have the same melody.

\begin{figure}[!t]
    \centering
    \includegraphics[width=0.95\linewidth]{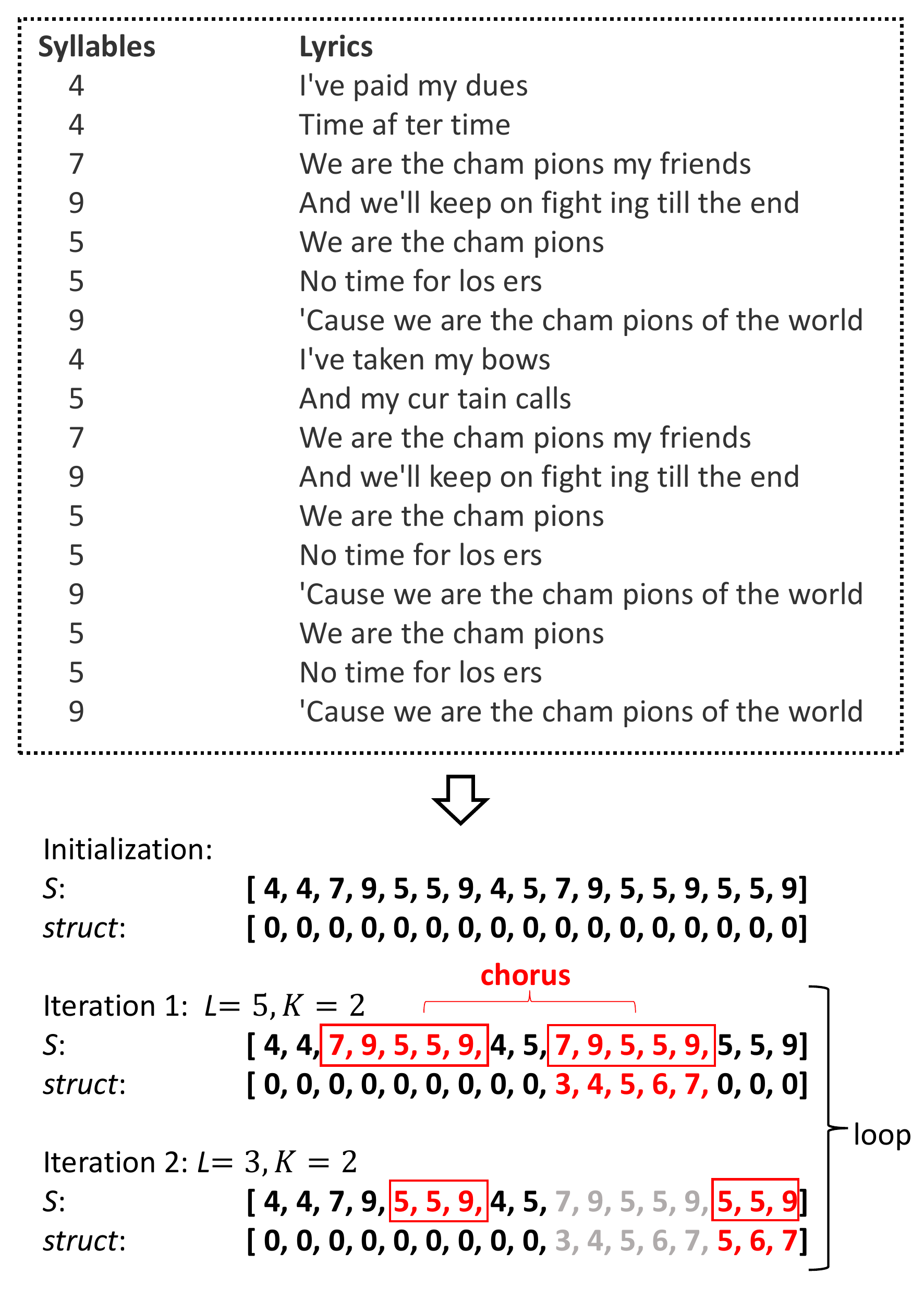}
    \caption{A case of Algorithm ~\ref{alg:lyric-recog}. Due to page limit and demonstration of algorithm properties, we simplify the original lyrics. We use red to highlight the operation in each step and grey to indicate these elements are skipped in string $S$.}
    \label{fig:lyrics}
\end{figure}

Now, the problem turns to find $R[L,K]$ in $S$. In each iteration, we only search $R[L,K]$ with the longest $L$ greedily. Because in the first iteration, the algorithm finds the $R[L,K]$ with the global longest $L$, we regard this $R[L,K]$ as chorus (In reality, the chorus of a song is often the longest segment that repeats). To record structure, we introduce an auxiliary array of length $n$ called $struct$. If the $X$-th lyric should share with melody from the $Y$-th lyric, then the $X$-th element in $struct$ is assigned as $Y$. Initially, all elements in $struct$ are `0' which means no sharing relationship. We introduce a searching granularity $g$ to control the minimum length of repetitive segments. When searched $R[L,K]$ with $L$ shorter than $g$ or $K$ less than 1, the algorithm stops. In ROC, we set $g$ as 2 by default. The algorithm details are shown in Algorithm ~\ref{alg:lyric-recog}. Figure ~\ref{fig:lyrics} is an intuitive illustration of the algorithm. We choose \textit{We Are the Champions}\footnote{\url{https://www.youtube.com/watch?v=04854XqcfCY}} by Queen. In the first iteration, the algorithm recognizes that the chorus is from `We are the champions my friends' to `Cause we are the champions of the world'. The second chorus shares melody with the first chorus. In the second iteration, the second chorus recognized in last iteration is skipped because the $struct$ value is non-zero. The loop stops because there are no more repetitive segments longer than $g$, which is 2 here. Lyrics corresponding to zero $struct$ values will retrieve melodies independently in retrieval and re-ranking stage.

\subsubsection{Retrieval and Re-ranking}
\label{sec:retrieval}
With extracted features, ROC does the following operations to assign the best matching melody to each lyric sentence: 

(1) If the $struct$ value of a lyric is `0', we use both features extracted from lyrics and user-designated chords to retrieve melody candidates from the database. In ~\ref{apx:4}, we visualize how to use features and designated chords to retrieve, for better understanding. Considering some characteristics in ~\cref{sec:2-1-1}, retrieved candidates are filtered with the following composition guidelines and music theories: 

\begin{itemize}
    \item The first note of a song should be in range of G3 to F4. This is to prevent an overly pitched verse followed with a further higher pitched chorus, making candidates rare.
    \item The first pitch in a melody fragment must be less than 8 semitones apart from the last pitch of the melody context  (concatenated melody fragments of previous lyrics). This is for the ease of singing.
    \item The tendency of the last pitch in the context should be satisfied as much as possible. This is to make the two melodies connect more naturally.
\end{itemize}

After filter, we concatenate each candidate with the melody context, and let the melody language model score these candidates. For the sake of diversity, we randomly select from top-k melody fragments as the final result.

(2) If the $struct$ value of a lyric is non-zero, we use the $struct$ value as the index to find which melody of lyric should the current lyric share with. For example, in Figure ~\ref{alg:lyric-recog}, when composing the second chorus, we directly reuse melodies of the first chorus.

(3) Specical cases. Because melody fragments have two bars at most, some long lyrics may retrieve no candidates. In this case, we split the lyrics into pieces and retrieve for each piece then concatenate. Besides, ROC supports one syllable aligning with multiple notes which appears with a certain probability. When it happens, ROC retrieves fragments having more notes than the number of syllables in the lyric and randomly decide which notes connect.

\subsubsection{Concatenation and Polish}
We concatenate retrieved fragments as the final composition result. During concatenation, we calculate a threshold: the average rest time between endpoint notes in adjacent bars. We ensure the interval before the newly selected bar should be approximate to the threshold, which stabilizes the rhythm of the song. For better melodic generation, we further polish the song if adjacent lyrics have the same number of syllables: some of their notes are randomly discarded and re-retrieved so that they have similar rather than same melodies.

\section{Experimental Settings}

\subsection{Dataset}
We use LMD-matched MIDI dataset~\cite{raffel2016learning}, which contains 45,129 MIDI data.  First, we separate tracks~\cite{guo2020variational} and extract melodies. Tonalities are normalized to “C major” or “A minor”. Ten percent of the data constitutes the validation set for training the melody language model. All data are used for constructing the database through the short melody fragment generation procedure, and there are 139,678 records in the database. As the test set, we select 20 English and Chinese songs respectively, and label their structures manually.

\subsection{Models}
The melody language model in ROC is a 4-layer decoder-only transformer~\cite{NIPS2017_3f5ee243}. Each layer has 4 attention heads, and 256 input/output dimension. We use Adam optimizer~\cite{kingma2014adam} with Adam $\beta$=$(0.9, 0.98)$. The initial learning rate is 0.0001. We apply early stop scheme with 20 epochs patience and select the best checkpoint by perplexity on the valid set. The model applies top-5 decoding scheme. 

With respect to baselines, we choose SongMASS~\cite{DBLP:journals/corr/abs-2012-05168} and TeleMelody~\cite{DBLP:journals/corr/abs-2109-09617} as representative of end-to-end models and non end-to-end models, respectively. Because the original SongMASS is trained with English lyrics, we follow ~\cite{DBLP:journals/corr/abs-2109-09617} to obtain a Chinese version. To evaluate the effectiveness of our lyric structure recognition algorithm, we also compare our algorithm with self-similarity matrix based methods ~\cite{fell-etal-2018-lyrics,watanabe-etal-2016-modeling} with pretrained embeddings GloVe~\cite{pennington2014glove} for English and CA8~\cite{li-etal-2018-analogical} for Chinese.

\begin{table*}[!t]
    \centering
    \small
    \begin{tabular}{@{}lcccccccccc@{}}
    \toprule
    \multirow{2}{*}{\textbf{Models}}& \multicolumn{4}{c}{\textbf{Objective}} & \multicolumn{5}{c}{\textbf{Subjective}}\\
\cmidrule(l{2pt}r{2pt}){2-5}\cmidrule(l{2pt}r{2pt}){6-10}
    &\textbf{Dist-1}& \textbf{Dist-2} &\textbf{Ent-1} &\textbf{Ent-2} &\textbf{Struc} &\textbf{Rhy} & \textbf{LMC} &\textbf{CLC} &\textbf{Melo}\\ \midrule
    SongMASS (EN)~\cite{DBLP:journals/corr/abs-2012-05168} & 0.62 & 4.32 & 2.18 & 3.83 & 2.80 & 2.80 & 2.60 & 3.00 & 3.30 \\
    TeleMelody (EN)~\cite{DBLP:journals/corr/abs-2109-09617} & 0.81 & 4.61 & 2.30 & 3.88 & 3.20 & 3.30 & 2.90 & 3.80 & 3.80\\
    \textbf{ROC} (EN)& \textbf{0.97} & \textbf{6.81} & \textbf{2.58}  & \textbf{4.40} & \textbf{4.50} & \textbf{4.00} & \textbf{4.20}  & \textbf{4.00} & \textbf{4.00} \\ \midrule
    SongMASS (ZH)~\cite{DBLP:journals/corr/abs-2012-05168} & 0.45 & 3.97 & 2.05 & 3.75 & 2.30 & 2.60 & 2.50 & 3.00 & 2.90\\
    TeleMelody (ZH)~\cite{DBLP:journals/corr/abs-2109-09617} & 0.75 & 4.54 & 2.32 & 3.85 & 3.20 & 3.50  &2.90 & 3.80 & 3.70\\
    \textbf{ROC} (ZH)& \textbf{0.91} & \textbf{6.60} & \textbf{2.57}  & \textbf{4.41} & \textbf{4.50} & \textbf{4.10} & \textbf{4.20}  & \textbf{4.00} & \textbf{4.10} \\  \bottomrule
    \end{tabular}
    \caption{Objective and subjective evaluation results on Chinese and English lyric-to-melody test set.}
    \label{tab:res}
\end{table*}

\subsection{Evaluation Metrics}

We conduct objective and subjective experiments. The lack of universal metrics is a notorious problem in melody generation area. In previous works such as TeleMelody~\cite{DBLP:journals/corr/abs-2109-09617}, researchers always measure the differences between generated songs and ground truth. However, we argue that pursuing the similarity with the ground truth cannot evaluate the creativity of the model. In this paper, we mainly rely on human evaluation (subjective metrics) and use objective metrics to only \textit{qualitatively} reflect generation quality. 

\textbf{Objective Metrics.} (1) Diversity (Dist-n)~\cite{li-etal-2016-diversity}: this metric is widely used in NLP fields to measure the diversity of generation, i.e., how many unique n-grams in generated songs. This metric can measure the quality of music to a certain extent because a song having few unique n-grams is very monotonous. (2) Entropy (Ent-n) ~\cite{10.5555/3326943.3327110}: Dist-n neglects the frequency difference of n-grams. As a complement, we also compute Entropy which reflects how evenly the n-gram distribution is for a given melody. (3) IoU (Intersection Over Union): we test the accuracy of lyric structure recognition by IoU. The ground truth are lyric structures in origin songs, which are annotated by humans. Because the accuracy of verse and chorus recognition are positively correlated, we only consider chorus accuracy.

\textbf{Subjective Metrics.} Objective metrics can only qualitatively reflect the generation quality and subjective metrics indicate that generation. Therefore, we recruit 10 evaluators having basic music knowledge to evaluate the performance of lyric-to-melody system from the following five aspects: (\uppercase\expandafter{\romannumeral1}) Structure (\textit{Struc}): how well the the melody structure matches lyric structure? Specifically, whether lyrics with similar rhythm patterns have similar melodies? (\uppercase\expandafter{\romannumeral2}) Rhythmic (\textit{Rhy}): is the rhythm of a song flexible? (\uppercase\expandafter{\romannumeral3}) Lyrics and melodies compatibility (\textit{LMC}): is lyric-melody feature alignment significant, e.g., when the lyrics enter the chorus, does the melody have a pitch lift or emotional intensity? Do similar lyrics share similar melodies? (\uppercase\expandafter{\romannumeral4}) Cadence and lyric ending compatibility (\textit{CLC}): whether cadences in the song sounds harmonic and whether there is an appropriate pause at the end of a lyric? (\uppercase\expandafter{\romannumeral5}) Melodic (\textit{Melo}): is the melody beautiful and attractive? In each aspect, evaluators can score from `1' for bad to `5' for good. 

Evaluators listen to all songs generated in experiments in a random order. To eliminate familiarity bias caused by evaluators having listened to original songs, we show their pseudo-lyrics (e.g., using symbols to occupy places where there are syllables). After scoring, they are told what true lyrics are and they focus on the structure to evaluate relevant aspects.

\section{Experimental Results}
\subsection{Main Results}

Table ~\ref{tab:res} shows results of objective and subjective judgement. In objective experiments, ROC outperforms baselines in each language. The comprehensive gains on all metrics demonstrate the effectiveness of our new paradigm for lyric-to-melody generation: 
(1) Higher diversity scores of ROC imply that there are more melodic motions, which prevents the melody being unattractive. More diverse melodies are more likely to promote the emotion expression. 
(2) Higher entropy scores indicate that diverse notes are distributed more evenly than those of baselines, that is, the attractiveness and the ability of better emotion expression are more likely to maintain from the start to the end. 

The above conclusions are confirmed in the subjective experiments, where ROC also outperforms baselines by a large margin in two languages:
(\uppercase\expandafter{\romannumeral1}) ROC achieves significant gains in \textit{Struc} thanks to lyric structure recognition and melody sharing scheme. In baselines, perhaps an implicit structural feature is captured during training, there are some weak structural patterns, but they are not as evident and neat as those in ROC. Also thanks to explicit structure features, we can distinguish chorus and verse which is an explicit activation for pitch range change or emotion expression promotion. 
(\uppercase\expandafter{\romannumeral2}) ROC has an improvement in \textit{Rhy} because of more flexible notes, e.g., durations vary much more often than those in baselines. This is because in ROC, fragments are short whereas baselines suffer from modeling longer-term dependency. 
(\uppercase\expandafter{\romannumeral3}) Due to the feature match between melody fragments and lyrics, we also beat baselines on \textit{LMC} by a large margin. 
(\uppercase\expandafter{\romannumeral4}) Because the pipeline of ROC includes pause and cadence polish, \textit{CLC} is ensured.
(\uppercase\expandafter{\romannumeral5}) Last but maybe the most important, ROC generates more melodic songs (highest \textit{Melo}). Better \textit{Struc}, \textit{Rhy}, \textit{LMC} and \textit{CLC} are also factors making songs more beautiful, improving \textit{Melo}.
Overall, both the objective and subjective evaluation results demonstrate that the new paradigm ROC outperforms conventional generation paradigm. The effect of each component in ROC will be discussed in detail in ~\cref{sec:abl}.

We test the accuracy of structure recognition algorithm. The average IoU of our algorithm reaches 0.77 with variance 0.09. By contrast, the average IoU of self-similarity matrix is 0.47, with variance 0.08, which is much worse. In fact, we find most of the errors in our algorithm come from pre-choruses that are classified as choruses, which is completely acceptable in terms of the sense of hearing. 

Our demos are in \url{https://ai-muzic.github.io/roc}. We also present a case study on structure recognition and a comparison with baselines in ~\ref{apx:1} and ~\ref{apx:3}. 

\subsection{Method Analyses}
\label{sec:abl}
To better study the effect of each component in ROC and explore properties of ROC more thoroughly, we analyze the impact of the structure recognition algorithm, model scores and composition guidelines in retrieval and re-ranking, and the size of database. Because of the slight performance difference in different languages, we report the average scores of two languages in below.

\textbf{Study on Structure Recognition.} We disable the lyric structure recognition and report results in Table ~\ref{tab:abla-no-share}. Because \textit{CLC} is guaranteed by polish operations in ROC, it is stable in this study and thus is omitted. Evaluators reflect that if we do not distinguish chorus and verse, the model will continue the song without an emotion activation or an explicit change of style so that melodies will be flat and less emotional, resulting in a smaller pitch range (lower Dist-1). Because lyric structure recognition is the foundation of the melody sharing scheme in ROC, without melody sharing, each sentence has its own unique melody, and thus Dist-2 increases. Overall, (w/o. recog.) impairs the generation quality by a large margin according to subjective evaluation because the rhythm is hurt and songs do not sound human-composed due to the lack of alignment between lyrics and melodies. Because the melody language model and guidelines ensure the basic quality and stability, the entropy scores maintain. 
This study reveals that aligning the structure of melodies to that of the lyrics is indispensable to high-quality lyric-to-melody generation. 

\begin{table}[!t]
    \resizebox{\linewidth}{!}{
    \centering
    \small
    \begin{tabular}{@{}lcccccccccc@{}}
    \toprule
    \multirow{2}{*}{\textbf{Models}}& \multicolumn{4}{c}{\textbf{Objective}} & \multicolumn{4}{c}{\textbf{Subjective}}\\
\cmidrule(l{2pt}r{2pt}){2-5}\cmidrule(l{2pt}r{2pt}){6-9}
  &\textbf{Dist-1}& \textbf{Dist-2} &\textbf{Ent-1} &\textbf{Ent-2} &\textbf{Struc} &\textbf{Rhy} & \textbf{LMC}  &\textbf{Melo}\\ \midrule
    ROC  & 0.94 & 6.71 & 2.58 & 4.41 & 4.50 & 4.10 & 4.20 & 4.10\\
    ROC w/o. recog. & 0.84 & 7.80 & 2.58 & 4.41 & 2.20 & 3.90 & 2.10  & 3.60\\ \bottomrule
    \end{tabular}}
    \caption{Study on lyric structure recognition.}
    \label{tab:abla-no-share}
\end{table}

\begin{table}[!t]
    \resizebox{\linewidth}{!}{
    \centering
    \small
    \begin{tabular}{@{}lccccccccc@{}}
    \toprule
    \multirow{2}{*}{\textbf{Models}}& \multicolumn{4}{c}{\textbf{Objective}} & \multicolumn{2}{c}{\textbf{Subjective}}\\
\cmidrule(l{2pt}r{2pt}){2-5}\cmidrule(l{2pt}r{2pt}){6-7}
     &\textbf{Dist-1}& \textbf{Dist-2} &\textbf{Ent-1} &\textbf{Ent-2}  &\textbf{Rhy}  &\textbf{Melo}\\ \midrule
       ROC   & 0.94 & 6.71 & 2.58 & 4.41 & 4.10 & 4.10  \\ 
    ROC w/o. model  & 0.69 & 3.57 & 2.41 & 3.62 & 3.80 & 3.70\\
    ROC w/o. guidelines & 0.91 & 9.07 & 2.76 & 4.79 & 4.20 & 3.60\\\bottomrule
    \end{tabular}}
    \caption{Study on re-ranking schemes.}
    \label{tab:abla-filter}
\end{table}

\begin{table}[!t]
    \resizebox{\linewidth}{!}{
    \centering
    	\small
    \begin{tabular}{@{}lccccccccc@{}}
    \toprule
    \multirow{2}{*}{\textbf{Database Size}}& \multicolumn{4}{c}{\textbf{Objective}} & \multicolumn{2}{c}{\textbf{Subjective}}\\
\cmidrule(l{2pt}r{2pt}){2-5}\cmidrule(l{2pt}r{2pt}){6-7}
     &\textbf{Dist-1}& \textbf{Dist-2} &\textbf{Ent-1} &\textbf{Ent-2} &\textbf{Rhy} &\textbf{Melo}\\ \midrule
    20\% & 0.93 & 6.54 & 2.49 & 4.28 & 3.80 & 3.90\\
    50\% & 0.94 & 6.67 & 2.57 & 4.34 & 3.80 & 4.00\\
    80\% & 0.96 & 6.60 & 2.56 & 4.35 & 4.00 & 4.10\\\midrule
    100\% & 0.94 & 6.71 & 2.58 & 4.41 & 4.10 & 4.10 \\
        \bottomrule
    \end{tabular}}
    \caption{Study on database size.}
    \label{tab:abla-size}
\end{table}

\textbf{Study on Model Scores and Composition Guidelines.}
We study the impact of model scores and composition guidelines in retrieval and re-ranking on the performance of ROC. We remove the melody language model and composition guidelines respectively. Table ~\ref{tab:abla-filter} shows experimental results. Because \textit{Struc}, \textit{CLC} and \textit{LMC} are unrelated to this study, their scores hardly change and thus are omitted.

With only composition guidelines, too many candidates remain, and thus there is a lot of randomness in the final determination. The melodies are so diverse that \textit{Rhy} increases a little. But too much diversity also decreases \textit{Melo}. 

When composition guidelines are removed, there are also too many candidates remaining for the melody language model to score, and thus the running speed is 70 times slower than that of ROC with only composition guidelines. Lower diversity and entropy indicate that the melody is monotonous, which can be confirmed by subjective metrics \textit{Rhy} and \textit{Melo}. 

Overall, when composition guidelines are removed, songs sound dull and the melody progression is not harmonic as before whereas when the melody language model scores are removed, the quality of different parts of a song varies because of randomness. This study reveals that the melody language model scores and composition guidelines complement each other in retrieval and re-ranking, which are both crucial to the quality and efficiency of ROC.

\textbf{Study on the Size of Database.}
Performance of ROC depends on the size of database. For example, if there is no melody that satisfies both the length requirements and chord progressions, ROC has to compromise, e.g., using the tonic chord as an alternative. Therefore, we study the effect of the size of database. We prune the database to 20\%, 50\%, and 80\% of the full size, respectively. Results are listed in Table ~\ref{tab:abla-size}. We have conclusions as below. 

First, as we expect, the running time and the database size are positively correlated. The running time increases from 3.99 seconds per song to 10.17 seconds per song when the size increases from 20\% to 100 \%. 
Second, because we remove data from the database randomly, the average quality and distribution of music fragments do not change, so \textit{Dist} and \textit{Ent} basically maintain. Because \textit{Struc}, \textit{LMC}, \textit{CLC} is not determined by the database size and these metrics do not change, they are omitted. To our surprise, we find that as long as the average quality of melody fragments is satisfying, the generation quality is stable even though only 20\% data remain. However, in practice, when we use 20\%-size database, sometimes there are no candidates with matching features.

Besides, we study the method of generating melody fragments in the database. If we generate a long melody and separate it into fragments instead of conducting short melody fragment generation procedure, all metrics drop by a lot because of low-qualify melody fragments in the database.

\section{Conclusion}
In this paper, we propose ROC, a new paradigm for lyric-to-melody generation, which divides the end-to-end generation into two stages: creation and re-creation. In creation stage, ROC generates a large amount of short music fragments and store them in a database indexed by key features including chords, tonality, structural information. In re-creation stage, ROC recreates melody by retrieving fragments according to key features extracted from each lyric and concatenate them based on melody language model scores and composition guidelines. ROC does not need paired lyric-melody data for training and better aligns features between lyrics and melodies. Both objective and subjective experimental results demonstrate the effectiveness of ROC. In the future, there is some research to be explored such as how to model one syllable aligning with multiple notes by neural networks, how to take accompany into consideration, or how can neural models help the lyric structure recognition. Moreover, we hope to apply the motivation of ROC to other NLP tasks like knowledge-grounded dialogue and story telling.

\bibliographystyle{named}
\bibliography{ijcai23}

\begin{thebibliography}{}

\bibitem[\protect\citeauthoryear{Bao \bgroup \em et al.\egroup
  }{2018}]{DBLP:journals/corr/abs-1809-04318}
Hangbo Bao, Shaohan Huang, Furu Wei, Lei Cui, Yu~Wu, Chuanqi Tan, Songhao Piao,
  and Ming Zhou.
\newblock Neural melody composition from lyrics.
\newblock {\em CoRR}, abs/1809.04318, 2018.

\bibitem[\protect\citeauthoryear{Deng}{2020}]{cnsenti}
Da~Deng.
\newblock Chinese sentiment analysis library, which supports counting the
  number of different emotional words in the text.
\newblock \url{https://github.com/hiDaDeng/cnsenti}, 2020.

\bibitem[\protect\citeauthoryear{Fell \bgroup \em et al.\egroup
  }{2018}]{fell-etal-2018-lyrics}
Michael Fell, Yaroslav Nechaev, Elena Cabrio, and Fabien Gandon.
\newblock Lyrics segmentation: Textual macrostructure detection using
  convolutions.
\newblock In {\em Proceedings of the 27th International Conference on
  Computational Linguistics}, pages 2044--2054, Santa Fe, New Mexico, USA,
  August 2018. Association for Computational Linguistics.

\bibitem[\protect\citeauthoryear{Fukayama \bgroup \em et al.\egroup
  }{2010}]{japan}
Satoru Fukayama, Kei Nakatsuma, Shinji Sako, Takuya Nishimoto, and Shigeki
  Sagayama.
\newblock Automatic song composition from the lyrics exploiting prosody of
  japanese language.
\newblock {\em Proceedings of the 7th Sound and Music Computing Conference, SMC
  2010}, 01 2010.

\bibitem[\protect\citeauthoryear{Guo \bgroup \em et al.\egroup
  }{2020}]{guo2020variational}
Rui Guo, Ivor Simpson, Thor Magnusson, Chris Kiefer, and Dorien Herremans.
\newblock A variational autoencoder for music generation controlled by tonal
  tension.
\newblock In {\em Joint Conference on AI Music Creativity (CSMC + MuMe)}, 2020.

\bibitem[\protect\citeauthoryear{Jayaram}{2018}]{pychorus}
Vivek Jayaram.
\newblock Pychorus: Python module for detecting musical choruses.
\newblock \url{https://github.com/vivjay30/pychorus}, 2018.

\bibitem[\protect\citeauthoryear{Ju \bgroup \em et al.\egroup
  }{2021}]{DBLP:journals/corr/abs-2109-09617}
Zeqian Ju, Peiling Lu, Xu~Tan, Rui Wang, Chen Zhang, Songruoyao Wu, Kejun
  Zhang, Xiangyang Li, Tao Qin, and Tie{-}Yan Liu.
\newblock Telemelody: Lyric-to-melody generation with a template-based
  two-stage method.
\newblock {\em CoRR}, abs/2109.09617, 2021.

\bibitem[\protect\citeauthoryear{Kingma and Ba}{2015}]{kingma2014adam}
Diederik~P Kingma and Jimmy Ba.
\newblock Adam: A method for stochastic optimization.
\newblock In {\em International Conference on Learning Representations (ICLR)},
  2015.

\bibitem[\protect\citeauthoryear{Li \bgroup \em et al.\egroup
  }{2016}]{li-etal-2016-diversity}
Jiwei Li, Michel Galley, Chris Brockett, Jianfeng Gao, and Bill Dolan.
\newblock A diversity-promoting objective function for neural conversation
  models.
\newblock In {\em Proceedings of the 2016 Conference of the North {A}merican
  Chapter of the Association for Computational Linguistics: Human Language
  Technologies}, pages 110--119, San Diego, California, June 2016. Association
  for Computational Linguistics.

\bibitem[\protect\citeauthoryear{Li \bgroup \em et al.\egroup
  }{2018}]{li-etal-2018-analogical}
Shen Li, Zhe Zhao, Renfen Hu, Wensi Li, Tao Liu, and Xiaoyong Du.
\newblock Analogical reasoning on {C}hinese morphological and semantic
  relations.
\newblock In {\em Proceedings of the 56th Annual Meeting of the Association for
  Computational Linguistics (Volume 2: Short Papers)}, pages 138--143,
  Melbourne, Australia, July 2018. Association for Computational Linguistics.

\bibitem[\protect\citeauthoryear{Li \bgroup \em et al.\egroup
  }{2020}]{li-etal-2020-rigid}
Piji Li, Haisong Zhang, Xiaojiang Liu, and Shuming Shi.
\newblock Rigid formats controlled text generation.
\newblock In {\em Proceedings of the 58th Annual Meeting of the Association for
  Computational Linguistics}, pages 742--751, Online, July 2020. Association
  for Computational Linguistics.

\bibitem[\protect\citeauthoryear{Liang \bgroup \em et al.\egroup
  }{2020}]{DBLP:journals/corr/abs-2010-08091}
Hongru Liang, Wenqiang Lei, Paul~Yaozhu Chan, Zhenglu Yang, Maosong Sun, and
  Tat{-}Seng Chua.
\newblock Pirhdy: Learning pitch-, rhythm-, and dynamics-aware embeddings for
  symbolic music.
\newblock {\em CoRR}, abs/2010.08091, 2020.

\bibitem[\protect\citeauthoryear{Long \bgroup \em et al.\egroup
  }{2013}]{Long2013TMusicAM}
Cheng Long, Raymond~Chi wing Wong, and Raymond Ka~Wai Sze.
\newblock T-music: A melody composer based on frequent pattern mining.
\newblock {\em 2013 IEEE 29th International Conference on Data Engineering
  (ICDE)}, pages 1332--1335, 2013.

\bibitem[\protect\citeauthoryear{Loria}{2020}]{textblob}
Steven Loria.
\newblock Textblob v0.16.0 simple, pythonic, text processing--sentiment
  analysis, part-of-speech tagging, noun phrase extraction, translation, and
  more.
\newblock \url{https://github.com/sloria/textblob}, 2020.

\bibitem[\protect\citeauthoryear{Ma \bgroup \em et al.\egroup
  }{2021}]{10.1145/3474085.3475502}
Xichu Ma, Ye~Wang, Min-Yen Kan, and Wee~Sun Lee.
\newblock {\em AI-Lyricist: Generating Music and Vocabulary Constrained
  Lyrics}, page 1002–1011.
\newblock Association for Computing Machinery, New York, NY, USA, 2021.

\bibitem[\protect\citeauthoryear{Magenta}{2020}]{viterbi}
Magenta.
\newblock A serializable note sequence representation and utilities.
\newblock \url{https://github.com/magenta/note-seq}, 2020.

\bibitem[\protect\citeauthoryear{Malmi \bgroup \em et al.\egroup
  }{2016}]{10.1145/2939672.2939679}
Eric Malmi, Pyry Takala, Hannu Toivonen, Tapani Raiko, and Aristides Gionis.
\newblock Dopelearning: A computational approach to rap lyrics generation.
\newblock In {\em Proceedings of the 22nd ACM SIGKDD International Conference
  on Knowledge Discovery and Data Mining}, KDD '16, page 195–204, New York,
  NY, USA, 2016. Association for Computing Machinery.

\bibitem[\protect\citeauthoryear{Pennington \bgroup \em et al.\egroup
  }{2014}]{pennington2014glove}
Jeffrey Pennington, Richard Socher, and Christopher~D. Manning.
\newblock Glove: Global vectors for word representation.
\newblock In {\em Empirical Methods in Natural Language Processing (EMNLP)},
  pages 1532--1543, 2014.

\bibitem[\protect\citeauthoryear{Raffel}{2016}]{raffel2016learning}
Colin Raffel.
\newblock {\em Learning-based methods for comparing sequences, with
  applications to audio-to-midi alignment and matching}.
\newblock Columbia University, 2016.

\bibitem[\protect\citeauthoryear{Sheng \bgroup \em et al.\egroup
  }{2020}]{DBLP:journals/corr/abs-2012-05168}
Zhonghao Sheng, Kaitao Song, Xu~Tan, Yi~Ren, Wei Ye, Shikun Zhang, and Tao Qin.
\newblock Songmass: Automatic song writing with pre-training and alignment
  constraint.
\newblock {\em CoRR}, abs/2012.05168, 2020.

\bibitem[\protect\citeauthoryear{Vaswani \bgroup \em et al.\egroup
  }{2017}]{NIPS2017_3f5ee243}
Ashish Vaswani, Noam Shazeer, Niki Parmar, Jakob Uszkoreit, Llion Jones,
  Aidan~N Gomez, \L~ukasz Kaiser, and Illia Polosukhin.
\newblock Attention is all you need.
\newblock In I.~Guyon, U.~V. Luxburg, S.~Bengio, H.~Wallach, R.~Fergus,
  S.~Vishwanathan, and R.~Garnett, editors, {\em Advances in Neural Information
  Processing Systems}, volume~30. Curran Associates, Inc., 2017.

\bibitem[\protect\citeauthoryear{Watanabe \bgroup \em et al.\egroup
  }{2016}]{watanabe-etal-2016-modeling}
Kento Watanabe, Yuichiroh Matsubayashi, Naho Orita, Naoaki Okazaki, Kentaro
  Inui, Satoru Fukayama, Tomoyasu Nakano, Jordan Smith, and Masataka Goto.
\newblock Modeling discourse segments in lyrics using repeated patterns.
\newblock In {\em Proceedings of {COLING} 2016, the 26th International
  Conference on Computational Linguistics: Technical Papers}, pages 1959--1969,
  Osaka, Japan, December 2016. The COLING 2016 Organizing Committee.

\bibitem[\protect\citeauthoryear{Wu \bgroup \em et al.\egroup
  }{2020}]{WU2020103303}
Jian Wu, Xiaoguang Liu, Xiaolin Hu, and Jun Zhu.
\newblock Popmnet: Generating structured pop music melodies using neural
  networks.
\newblock {\em Artificial Intelligence}, 286:103303, 2020.

\bibitem[\protect\citeauthoryear{Xue \bgroup \em et al.\egroup
  }{2021}]{DBLP:journals/corr/abs-2107-01875}
Lanqing Xue, Kaitao Song, Duocai Wu, Xu~Tan, Nevin~L. Zhang, Tao Qin,
  Wei{-}Qiang Zhang, and Tie{-}Yan Liu.
\newblock Deeprapper: Neural rap generation with rhyme and rhythm modeling.
\newblock {\em CoRR}, abs/2107.01875, 2021.

\bibitem[\protect\citeauthoryear{Yu \bgroup \em et al.\egroup
  }{2021}]{10.1145/3424116}
Yi~Yu, Abhishek Srivastava, and Simon Canales.
\newblock Conditional lstm-gan for melody generation from lyrics.
\newblock {\em ACM Trans. Multimedia Comput. Commun. Appl.}, 17(1), apr 2021.

\bibitem[\protect\citeauthoryear{Zhang \bgroup \em et al.\egroup
  }{2018}]{10.5555/3326943.3327110}
Yizhe Zhang, Michel Galley, Jianfeng Gao, Zhe Gan, Xiujun Li, Chris Brockett,
  and Bill Dolan.
\newblock Generating informative and diverse conversational responses via
  adversarial information maximization.
\newblock In {\em Proceedings of the 32nd International Conference on Neural
  Information Processing Systems}, NIPS'18, page 1815–1825, Red Hook, NY,
  USA, 2018. Curran Associates Inc.

\bibitem[\protect\citeauthoryear{Zhu \bgroup \em et al.\egroup
  }{2018}]{xiaoice}
Hongyuan Zhu, Qi~Liu, Nicholas~Jing Yuan, Chuan Qin, Jiawei Li, Kun Zhang,
  Guang Zhou, Furu Wei, Yuanchun Xu, and Enhong Chen.
\newblock Xiaoice band: A melody and arrangement generation framework for pop
  music.
\newblock In {\em Proceedings of the 24th ACM SIGKDD International Conference
  on Knowledge Discovery and Data Mining}, page 2837–2846, New York, NY, USA,
  2018. Association for Computing Machinery.

\end{thebibliography}

\appendix
\clearpage

\section{Case Study}
\label{apx:1}
\begin{figure}[h]
\centering 
\subfigure[Lyrics]{ 
\begin{minipage}{0.8\linewidth}
\centering   
\includegraphics[width=\linewidth]{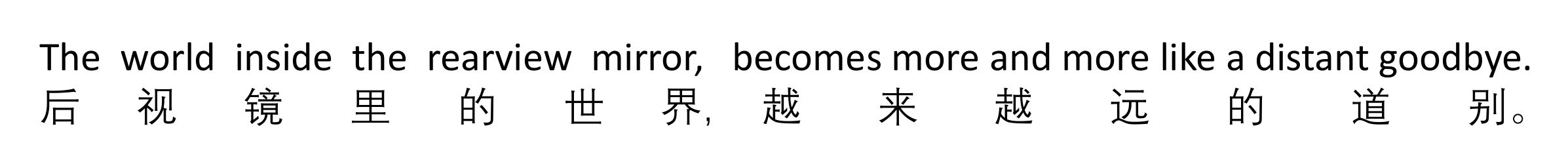}
\end{minipage}
}

\subfigure[Orginal version]{ 
\begin{minipage}{\linewidth}
\centering   
\includegraphics[width=\linewidth]{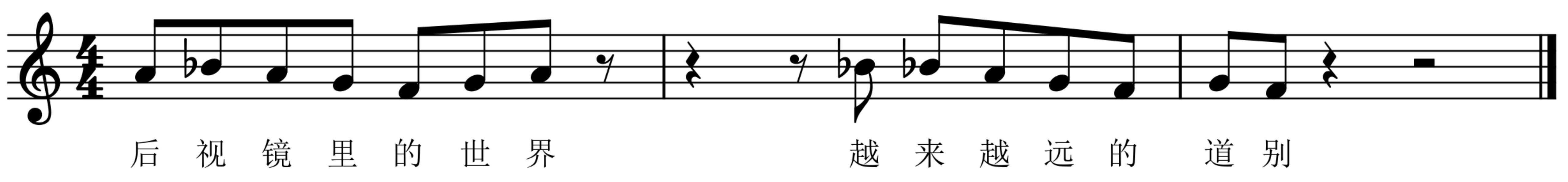}
\end{minipage}
}

\subfigure[TeleMelody]{ 
\begin{minipage}{\linewidth}
\centering   
\includegraphics[width=\linewidth]{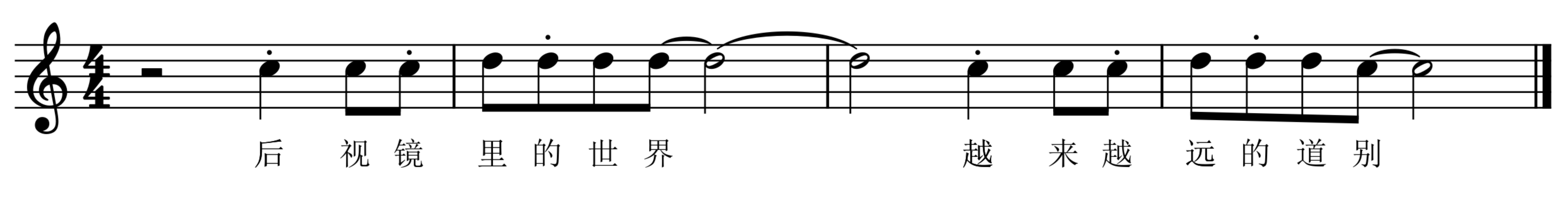}
\end{minipage}
}

\subfigure[ROC]{   
\begin{minipage}{\linewidth}
\centering    
\includegraphics[width=\linewidth]{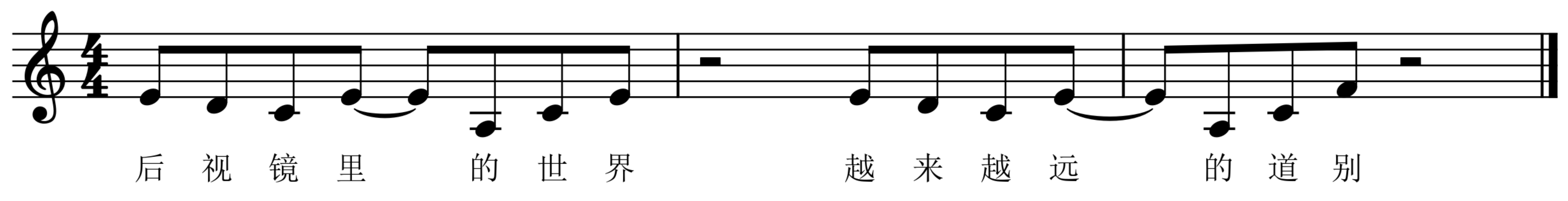} 
\end{minipage}
}
\caption{A case of melodic and pitch-suitable generation. Melody from ROC is more diverse and in a proper pitch range. In ROC, rhythm alignment between lyrics and melodies is guaranteed.}
\label{fig:case2}
\end{figure}

To highlight that ROC is not limited by any specific language, we choose a Chinese case in Figure ~\ref{fig:case2} to demonstrate the generation quality. Lyrics are from the beginning of a famous Chinese pop song \textit{All the Way North}\footnote{\url{https://www.youtube.com/watch?v=OoM-97XZVLE}} by Jay Chou. In this case, we compare ROC with the origin song and the best-performing baseline, TeleMelody.

$\bullet$ TeleMeldoy generates flat melodies while the output from ROC is more varied and cadenced, which sounds more melodic. The diversity in ROC is similar to that in the origin song. 

$\bullet$ These are the first two lyrics in the song, so melodies of TeleMelody are overly pitched, where some notes reach D5 which is difficult even for some professional singers. By contrast, pitches in ROC are appropriate.

$\bullet$ In this case, because the adjacent lyric sentences have the same number of syllables (7 Chinese characters), ROC shares the first 6 notes between two sentences and modifies the last note of the second sentence. This causes the last characters in two sentences have the same melody duration which sounds more regular and is better than TeleMelody.

More demos are presented as videos and MP3 files in supplementary files.

\section{Data Structure in Database}
\label{apx:2}
\begin{figure*}[h!]
\centering   
\includegraphics[width=0.9\linewidth]{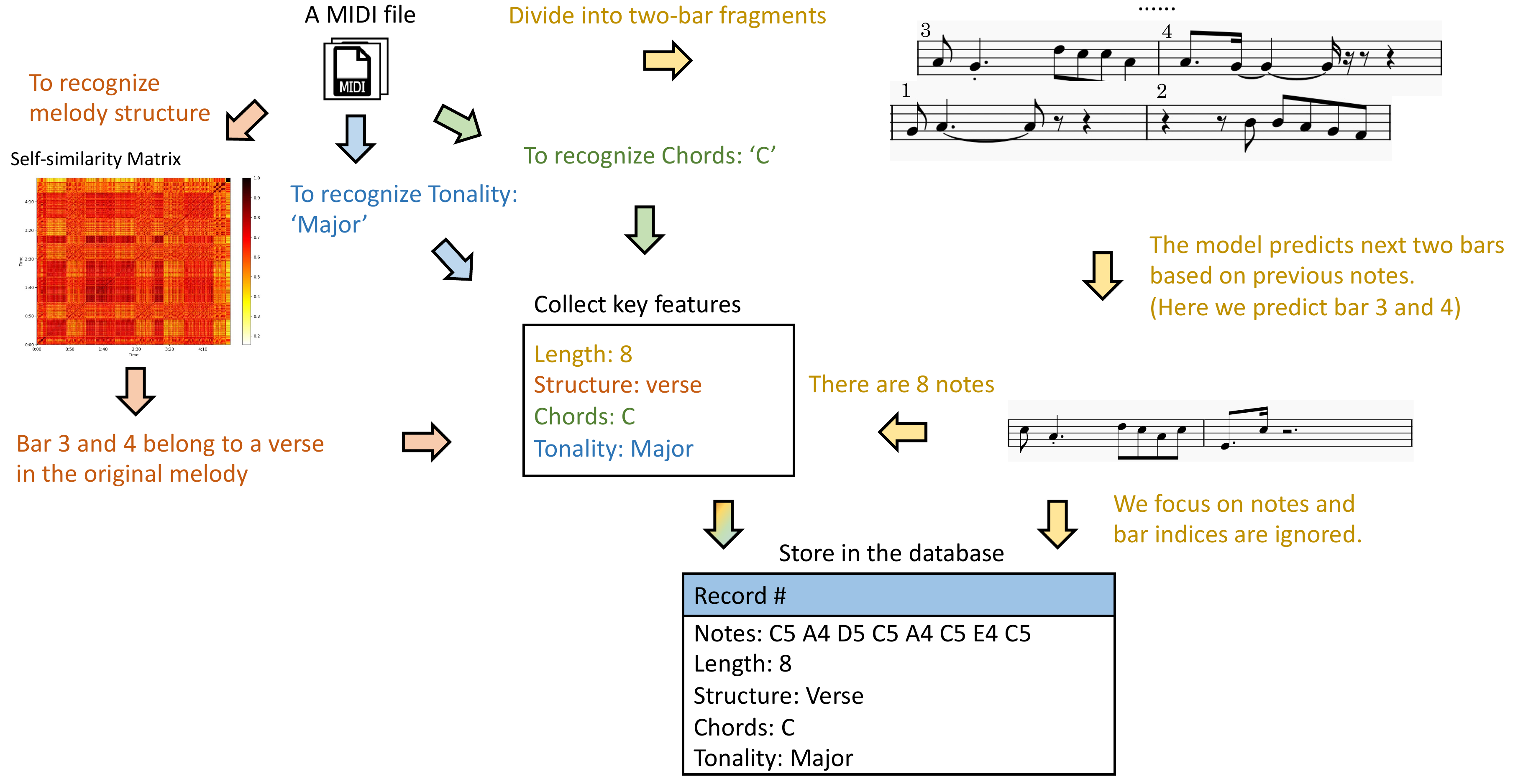}
\caption{An example of data structure.}
\label{fig:keyfeatures}
\end{figure*}

Here is a detailed supplement to the \textit{Melody Feature Extract}, Section 3.1 of the paper. We illustrate the process of storing a record in the database. Given a MIDI file, we divide it into two-bars fragments. The melody language model predicts the next two bars based on previous melody contexts. In Figure ~\ref{fig:keyfeatures}, we predict bar $\#$3 and $\#$4 based on bar $\#$1 and $\#$2 for example. As mentioned in the paper, we ignore the bar indices and focus on notes. We translate notes to MIDI representations. 

Meanwhile, we extract key features. For example, we construct a self-similarity matrix based on audio frequency of the file. Through the matrix, we can estimate from when to when is a chorus or a verse. Suppose bar $\#$3 and $\#$4 are verses in the original melody, and we regard the predicted bar $\#$3 and $\#$4 as a part of verses, too. We store notes and its key features in the database.

\section{Lyrics Structure Recognition}
\label{apx:3}
\begin{figure*}[h]
\centering   
\includegraphics[width=0.9\linewidth]{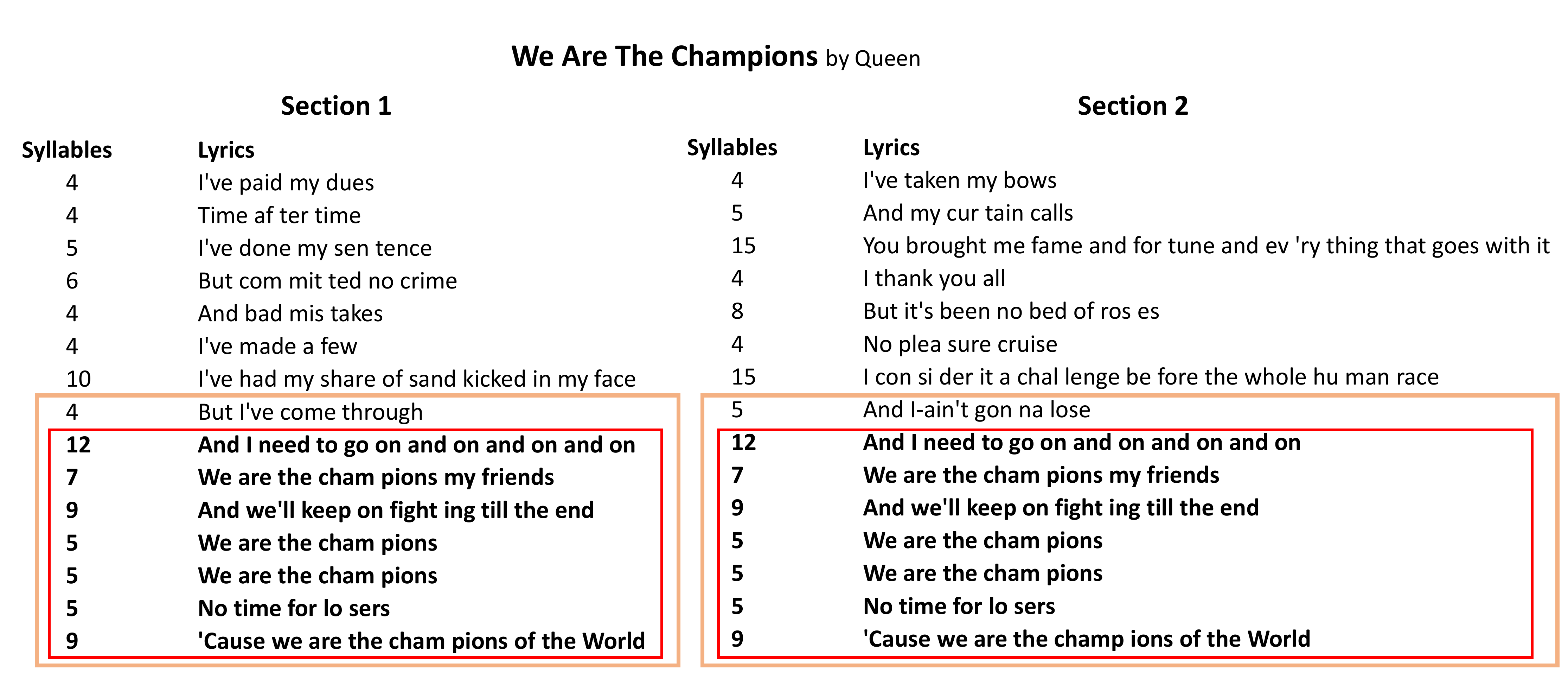}
\caption{Extracted lyric structure of \textit{We Are The Champions}.}
\label{fig:case_structure}
\end{figure*}

We visualize the recognized structure of the full version of \textit{We Are The Champions} in Figure ~\ref{fig:case_structure}. Choruses identified by our algorithm are in orange squares and ground truths are in the red squares. Because two verses have different rhythms, only choruses share melody in this case. The IoU of this case is 0.85 (12/14). Our algorithm regards the pre-chorus as chorus which is acceptable from an appreciative standpoint.

\section{Details on Re-creation Stage}
\label{apx:4}
\begin{figure*}[h!]
\centering   
\includegraphics[width=0.9\linewidth]{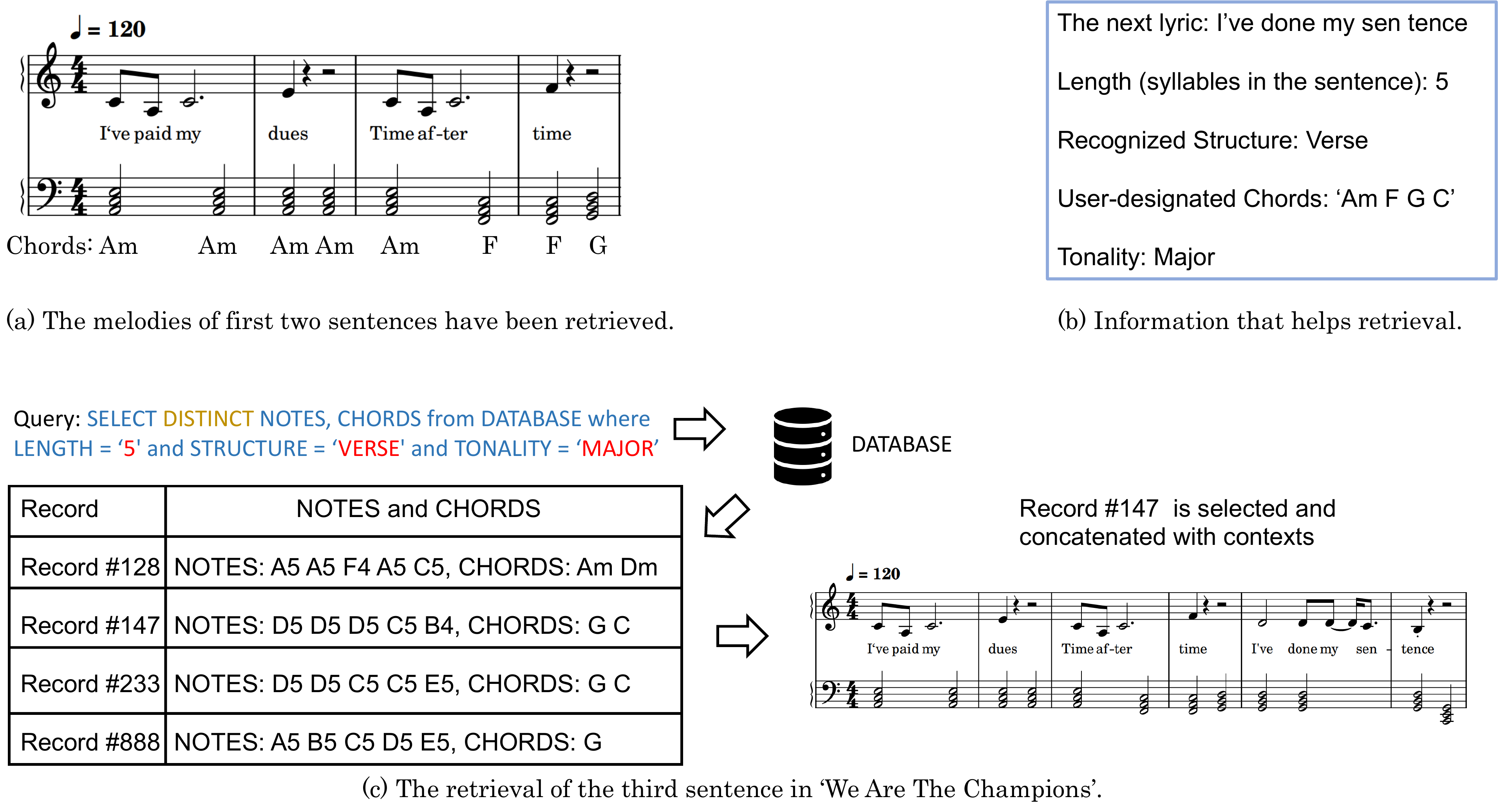}
\caption{An example illustrating how retrieval process works.}
\label{fig:recreate}
\end{figure*}

Suppose we are using the lyrics of `We Are The Champions' to compose and the melodies of the first two sentences have been decided. Based on key features of the thrid lyric sentence, we retrieve 3 records as shown in Figure~\ref{fig:recreate}(c).

According to the user-designated chords, the next melody fragments should still start with chord `G' or change to chord `C' because the last chord in the context is `G'. Therefore, we generate a regular expression like:

\^{} G( G)*( C)+( Am)+( F)*\$ $\vert$ \^{} G( G)*( C)+( Am)*\$ $\vert$ \^{}G( G)*( C)*\$ $\vert$ \^{} G( G)*\$

Given the regular expression, we filter retrieved records. We design the regular expression like this because alternations in regular expressions are evaluated from left to right, the record with the most various chords will be selected, if there's any. In this case, both record $\#$147, $\#$233 and $\#$888 start with chord `G' that satisfies user-designated chords after concatenation. Because record $\#$147 and $\#$233 has more various chords, they have higher priority to be selected while other records are discarded.

Next, we filter records with composition guidelines as Line 366 in the main paper mentioned. Suppose both record $\#$147 and $\#$233 are still kept. Then, we concatenate record $\#$147 and $\#$233 respectively with the melody contexts, i.e., the melody that has been decided. We use the melody language model to score the coherence of concatenated melodies. According to the model scores, we select the record $\#$147 as the final result.

\end{document}